\documentclass[debug,overfull]{epl}

\usepackage{amssymb}
\usepackage{subeqn}

\newcommand{\proj}{\mathcal{P}}
\newcommand{\ham}{\mathcal{H}}
\newcommand{\cdag}{c^\dagger}
\newcommand{\half}{\frac{1}{2}}
\newcommand{\splus}{S^+}
\newcommand{\sminus}{S^-}
\newcommand{\up}{\uparrow}
\newcommand{\down}{\downarrow}
\newcommand{\Up}{\Uparrow}
\newcommand{\Down}{\Downarrow}
\newcommand{\zero}{0}
\newcommand{\rbra}{\rfloor}
\newcommand{\lbra}{\lfloor}
\newcommand{\hole}{\mathbf{.}}

\title{Hole on a stripe in a spinless fermion model}
\author{U.~Hizi\thanks{E-mail: \email{uh22@cornell.edu}} \and C.~L.~Henley }
\institute{Laboratory of Atomic and Solid State Physics, 
Cornell University, Ithaca, NY, USA, 14853-2501 }

\pacs{71.10.Fd}{Lattice fermion models}
\pacs{71.10.Pm}{Fermions in reduced dimensions}

\begin{document}
\maketitle
\begin{abstract} 
In the spinless fermion model on a square lattice with 
infinite nearest-neighbor repulsion, holes doped into
the half-filled ordered state form stripes which,
at low doping, are stable against phase separation
into an ordered state and a hole-rich metal.
Here we consider transport of additional holes
along these stripes. 
The motion of a single hole on a stripe
is mapped to a one-dimensional problem,
a variational wavefunction 
is constructed and
the energy spectrum is calculated and 
compared to energies obtained by exact diagonalization.
\end{abstract}
\section{\label{sec:intro}Introduction}
A spinless square lattice fermion model with nearest-neighbor
repulsion, we suggest, may be a fruitful and comparatively 
tractable analog of the Hubbard model that retains many of 
the its properties.
In particular, in the limit of infinite repulsion, near half-filling, 
the equilibrium state appears to be an array 
of charged, antiphase domain wall stripes that are 
stabilized by kinetic energy~\cite{clh_stripe,ng_stripe}.
  This is reminiscent of a state of charged stripes that has been observed in
cuprates \cite{tranquata} and has been discussed in the spinfull
Hubbard and $t$-$J$ 
models~\cite{stripes}. 
Stripes in the spinless fermion model have been studied by 
exact diagonalization (ED) in  ref.~\cite{ng_stripe}.
Here we are interested in finding to what extent a hole, moving along a stripe,
 is decoupled from the state of the stripe, and whether or not
the hole tends to bind to a ``kink'' on the stripe.
To this end we extend an exact mapping of a stripe to one dimension,
 introduced in refs.~\cite{clh_stripe,ng_stripe}.
Using this mapping, we construct a variational wavefunction and calculate the
energy spectrum.

We do not know of an existing experimental system that realizes our model,
but recent progress in the manipulation of ultracold bosonic atoms on an 
optical 
lattice~\cite{greiner}, and in cooling fermionic atoms below degeneracy 
temperature~\cite{ohara}, leads us to expect that such a model may be realized 
experimentally in the near future. 
A corresponding Hard-core boson model with large nearest neighbor repulsion 
can be realized, on a triangular lattice, in adsorption of $\mathrm{^4He}$ to 
graphite sheets~\cite{green_prl}. 
The phase diagram of a boson model with finite nearest neighbor and
next nearest neighbor repulsion was studied in ref.~\cite{hebert}.

\section{\label{sec:model}Model}
We consider spinless fermions on a square lattice with Hamiltonian
\begin{equation}
\ham=-t \sum_{ \langle i j \rangle} (\cdag_i c_j + \cdag_j c_i)+
 V \sum_{\langle i j \rangle}  \cdag_i c_i  \cdag_j c_j\,,
\label{eq:ham}
\end{equation}
where $\cdag_i$ and $c_i$ are creation and annihilation operators 
at site $i$, respectively, 
and $\langle i j \rangle$ means nearest neighbors.
In this paper, we consider only the limit of zero temperature and
$V/|t|=\infty$, so that neighboring fermions are forbidden, and
$t \equiv 1$ is the only energy scale. The maximum allowed 
filling fraction is $n=1/2$, where there are two possible 
``checkerboard'' states~\cite{mila}.
To make a well-defined situation, we assume a finite system with
periodic boundary conditions having dimensions
$L_x\! \times \!L_y$ where $L_x$ is even and $L_y$ is odd.
This forces an odd number of domain walls (which must be stripes) running
in the $x$ direction. These domain walls are composed of $1/2$ of a hole
per column and we call them \emph{stripes}.
This model has been studied by ED for spinless fermions and hard-core bosons, 
and the stripes were shown to be stable against phase
separation, for fermions~\cite{clh_stripe,ng_stripe}. 
When the number of particles is $L_x(L_y\!-\!1)/2$, only
one stripe is allowed. If we remove a few more particles 
(less than $L_x$), it is energetically favorable for the holes to attach
 to the stripe, since holes off the stripe can only form confined 
droplets\cite{mila}. From here on, we reserve the term ``hole''
for additional holes beyond those needed to create a stripe.
In the case of a single undoped stripe,
the boson and fermion models
possess the same energy spectrum~\cite{ng_stripe}. 

On an undoped stripe, hops of particles 
are equivalent to stripe fluctuations (see fig.~\ref{fig:fluct}).
If we define the stripe height $y(x)$ to be the mean of the $y$ coordinates 
of the first particles  above and below the stripe, in column $x$,
we can map the up and down steps of the stripe height
to ``spins'' in one dimension by defining 
$ s(x) = \half [y(x)-y(x-1)]  $, taking values $\pm 1/2$.
The corresponding Hamiltonian is the one-dimensional spin-$1/2$ 
XY Hamiltonian
\begin{equation}
\ham_{ex}=-\sum_{i} \ham^i_{ex}\,,\qquad \mbox{where} \qquad
\ham^i_{ex}\ = \splus_i \sminus_{i+1} + \sminus_i \splus_{i+1} \,,
\label{eq:hxx}
\end{equation}
where $\splus_i$, $\sminus_i$ are spin raising and lowering operators.
This model can be solved by the Jordan-Wigner transformation, 
in which we replace each up spin by a spinless fermion, and each down spin
by an empty space. The Hamiltonian is replaced by a non-interacting 
hopping Hamiltonian for the fermions.
Thus, the ground state of a horizontal, undoped stripe is 
equivalent to that of a (one-dimensional) half-filled sea of 
free (spinless) fermions with dispersion $\epsilon(k) = -2 \cos k$.
From this, the ground state energy, for large $L_x$, 
 is $\epsilon_0^{(L_x)}=-2 L_x /\pi$.
This implies the chemical potential of the stripe 
is $\mu_{stripe} = -4/\pi$,
since half of a particle is removed per unit length of a stripe.
In a more general case, we may force a stripe with some overall tilt
by replacing our rectangular boundary conditions with 
$(L_x,b)\!\times\!(0,L_y)$, for $L_x\!+\!b$ even.
In that case the spin chain has total spin $b$, 
 mapping to 1D fermions results in $N\! =\! (L_x\!+\!b)/2$ fermions and the
 ground state energy is  $\epsilon_0^{(L_x,N)}=-2 L_x \sin(N \pi/L) /\pi$~
\cite{ng_stripe,flux_footnote}. 
{Note that a diagonal stripe ($b=L_x$) 
has zero energy, as there are no allowed hops.

\section{One hole on a stripe} \label{sec:hole}
If an additional hole is added to an a stripe, the energy is lowered
due to the hopping energy of the hole along the stripe. We define the
ground state energy difference $\Delta\equiv E_{hole} - E_{stripe}$,
where $E_{hole}$, $E_{stripe}=\epsilon_0^{(L_x)}$ are ground state energies 
of a stripe with a hole and an undoped stripe, respectively.
For stability of a stripe array state, we must have 
$\Delta > \mu_\mathrm{stripe}$;
otherwise, doping would add holes to existing stripes,
probably forming a phase-separated droplet.
To obtain a variational bound for $\Delta$, we will make (below) an 
approximation by using a subset of the Hilbert space.

Once a single hole is added to column $x$ of the
stripe, the stripe height difference 
in the two columns adjacent to the hole can take values of $0$ or $\pm 2$
rather than $\pm 1$, i.e. 
 $s(x),s(x\!+\!1) \in \{0,\pm 1\}$ (see, for example, 
figs.~\ref{fig:strand},~\ref{fig:main_path}).
Thus, we get a pair of  ``spin-$1$'' impurities on both sides
 of the hole, at positions $x$ and $x\!+\!1$ on the spin chain. 
This looks like a Kondo model with two mobile impurities, that are bound
together. 
Similar systems, with $t$-$J$ Hamiltonian and mobile spin-$1/2$ impurities 
in a spin-$1$ chain, have been the subject of theoretical study, 
as a model for the charge transfer insulator
$\mathrm{Y}_{2-x}\mathrm{Ca}_x \mathrm{BaNiO}_5$~\cite{spin1}.
However, our model is quite different from those,
because the hopping matrix is quite elaborate, 
reflecting the allowed moves in the original stripe. 

 In order for our variational scheme to work, we need to separate
the  ``hole degrees of freedom'' from the ``stripe degrees of freedom''.
To make this separation apparent, we find it more convenient 
to introduce a different, equivalent notation, in which the stripe 
and the hole is represented by a one dimensional spin-$1/2$ chain of 
length $L'_x \equiv L_x - 2$, with three additional ``particles'':
a hole (represented by a dot), that marks the position of the hole in between
two spins, and two ``brackets'' (right and left), 
each taking the place of a spin.
If $s(x\!+\!1)=1$ ($-1$), then the right bracket takes the places of the first 
down (up) spin to the right of the hole, and $s(x\!+\!1)$ is set to $1/2$ 
($-1/2$). If $s(x\!+\!1)=0$, the right bracket is placed in
position $x\!+\!1$. Similarly for the left bracket and $s(x)$.
Note that the total spin is preserved by this mapping.
For example, if we use a  double arrow to denote $s(x)=\pm 1$,
\begin{equation}
\begin{array}{ccccccccccccrccllllllllll}
\down  &\up  &\down  &\down  &\down  &\Down 
 & \Up  &\down  &\down  &\up  & \quad\Rightarrow \quad &
\down  &\!\!\!\lbra  \down  &\!\down  &\down  &\down\!~  \hole &
 \!\!\!\!\!\up  \rbra  &\!\!\!\down  &\!\!\!\up \\
\down  &\up  &\down  &\down  &\down  &\zero 
 &\Down  &\down  &\up  &\up  & \quad\Rightarrow \quad &
\down  &\!\!\!\up  &\!\down  &\down  &\down  \!\lbra  \hole &
  \!\!\!\!\!\down  &\!\!\!\down  \rbra   &\!\!\!\up \\
\up  &\down  &\up  &\up  &\up  &\Up
 &\Up  &\up  &\up  &\down  & \quad\Rightarrow \quad &
\up  &\!\!\!\lbra  \up  &\!\up  &\up  &\up \! ~\hole& 
  \!\!\!\!\!\up  &\!\!\!\up  &\!\!\!\up   \rbra
\end{array}
\label{eq:map}
\end{equation}
If stripe fluctuations occur in the vicinity of the hole, it might
become \emph{stranded}, i.e. isolated from the stripe, 
as in fig.~\ref{fig:strand}. 
However, we observe that in the ED ground state of ref.~\cite{ng_stripe},
the probability for this to occur is negligible 
(about $0.03$ for an untilted stripe and less than $0.01$ for a tilted one).
 Thus, in the following 
discussion we suppress hops that lead to the stranded state.
Under this assumption, the spin chain is free to fluctuate only outside
the brackets. Each of the three additional ``particles'' can hop 
by one step freely, as long as all of the
spins between the hole and each of the brackets are in the same direction,
 and as long as their order is preserved (i.e. they do not hop across 
each other). This mapping can be shown to be exact.

\begin{figure}
\threefigures[scale=0.3]{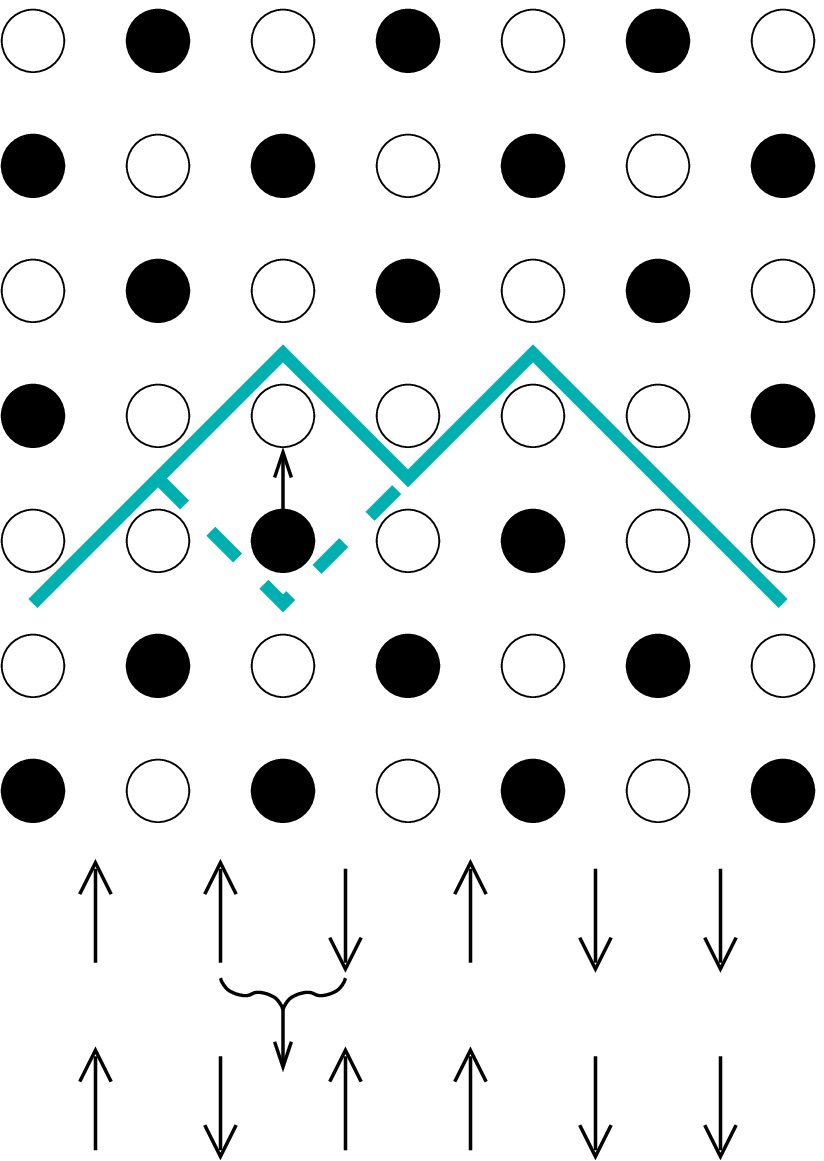}{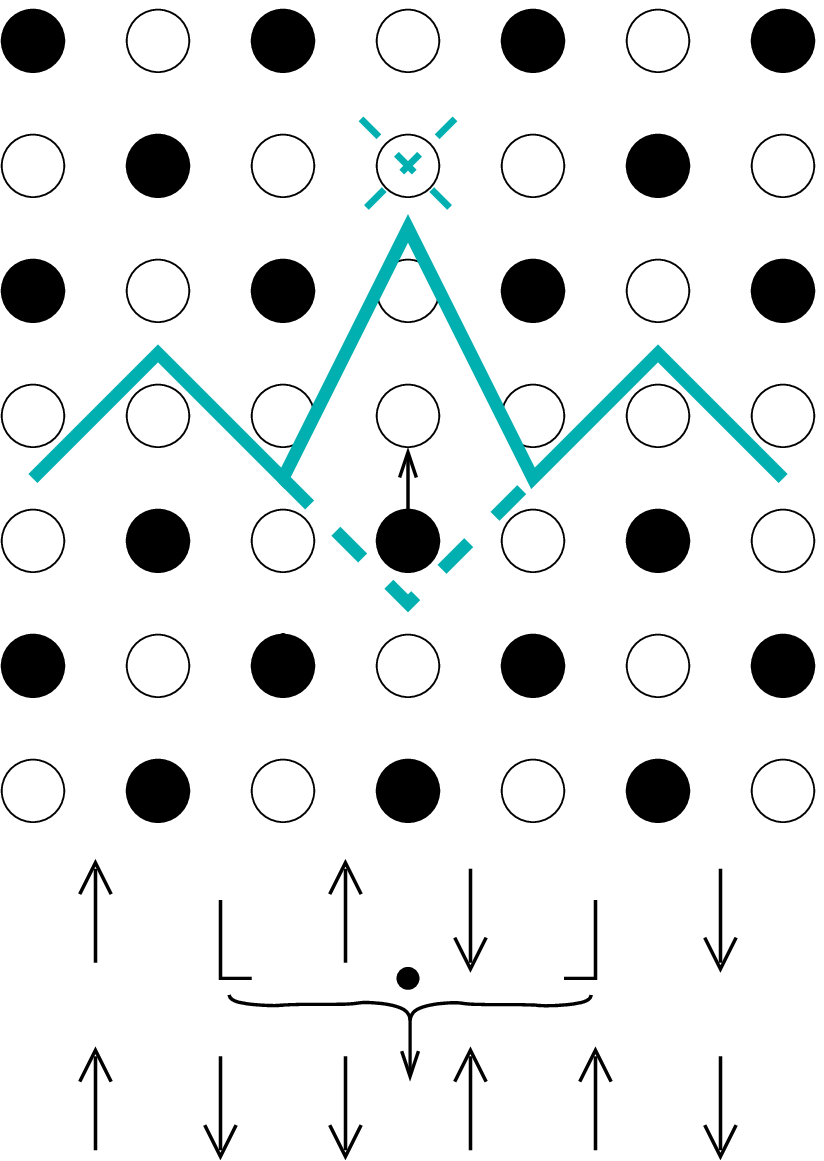}{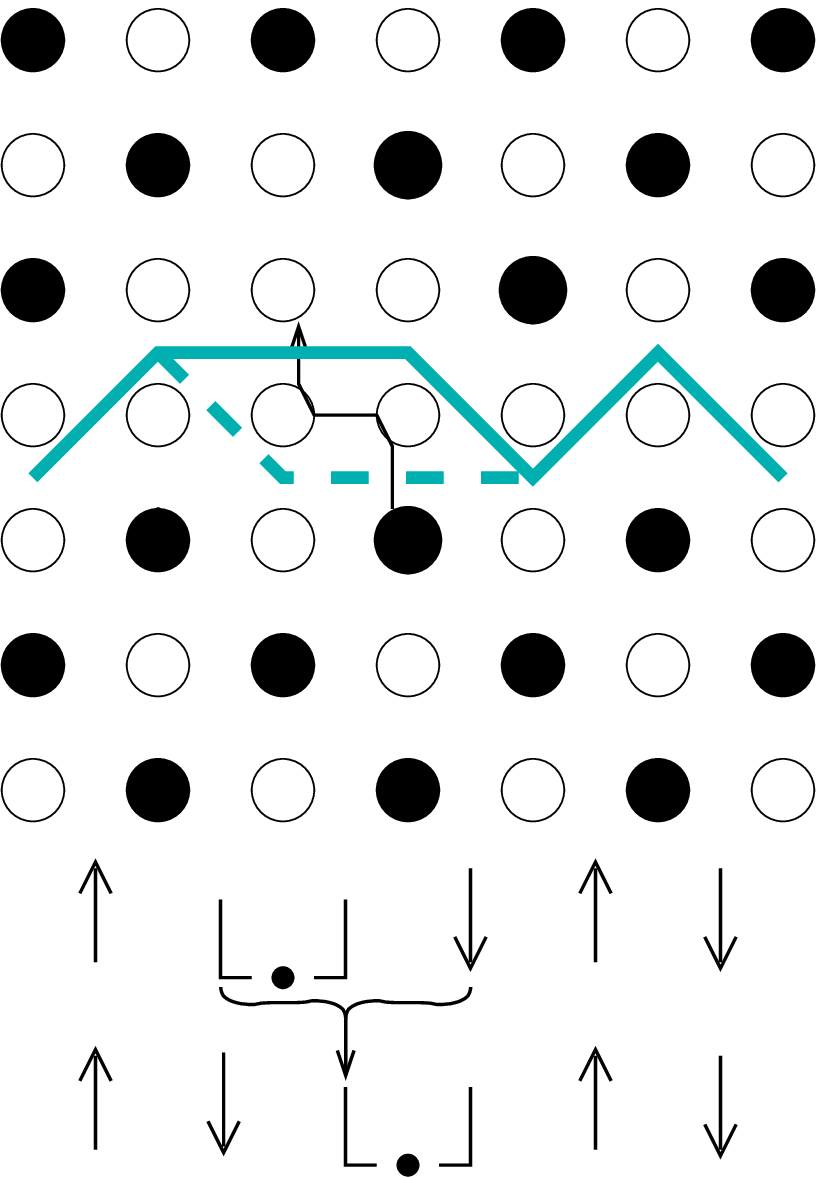}
\caption{Particle hops (stripe fluctuations)
 in an undoped stripe are equivalent to spin exchange.}
\label{fig:fluct}
\caption{Hole stranding: Stripe fluctuation across the hole 
can leave the hole away from the stripe, with no possible
 allowed hops. We ignore these moves our variational model.}
\label{fig:strand}
\caption{Three successive moves that shift the hole by one position 
to the right. The sequence shown here is a \emph{main path} sequence 
that can occur regardless of the configuration of the stripe.}
\label{fig:main_path}
\end{figure}

The Hamiltonian for the hole on the stripe in this model can now be 
broken up  $\ham_\mathrm{hole}=\ham_h + \ham_r + \ham_l + \ham'_{sp}$,
where $\ham_h$, $\ham_r$, $\ham_l$ are the hopping terms for the hole, right bracket and left bracket, respectively and $\ham_{sp}'$ is the spin
 exchange Hamiltonian acting only outside the brackets.
 The Hamiltonian $\ham_{hole}$ acts on the Hilbert space 
of ``allowed states'', i.e., all states in which the spins between each of the 
brackets and the hole are in the same direction.
 We denote the state of the hole and two brackets by $|x,l,r\rangle$,
where the hole is between the spins at 
positions $x$ and $x\!+\!1$, there are $l$ spins (sites $[x\!-\!l+\!1,x]$) 
between the 
hole and left bracket, and $r$ spins between the hole and the right bracket 
(sites $[x\!+\!1,x\!+\!r]$).
 For example, in the three examples in 
(\ref{eq:map}) above, if we number
the spins starting from $1$ at the left,  the hole+bracket states are 
$|5, 4, 1 \rangle$, $|5, 0, 2 \rangle$, $|5, 4, 3 \rangle$,
respectively.
Using this notation
\begin{subequations} \label{eq:ham_parts}
\begin{eqnarray}
\ham_h&=&
  - \sum_{x,l,r} \delta_{r>0}
\left( \mathcal{P}_x^{SS} + \delta_{l,0} (1-\mathcal{P}_x^{SS}) \right)
|x,l,r\rangle \langle x+1,l+1,r-1 | 
+ h.c. \,,
\label{eq:hh} \\
\ham_r&=&
- \sum_{x,l,r} (\mathcal{P}_{x+r}^{SS} + \delta_{r,0}(1-\mathcal{P}_{x+r}^{SS}))
 |x,l,r + 1\rangle \langle x,l,r|
+ h.c.\,,
\label{eq:hr} \\
\ham_l&=&
- \sum_{x,l,r} (\mathcal{P}_{x-l}^{SS} + \delta_{l,0}(1-\mathcal{P}_{x-l}^{SS}))
 |x,l + 1,r\rangle \langle x,l,r | 
+ h.c.\,,
\label{eq:hl} \\
\ham'_{sp} &=& -
\sum_{x l r} | x, l, r \rangle \langle x,l,r |
\left ({ \sum_i}' \ham_{ex}^i \right) \,,
\label{eq:hex'}
\end{eqnarray}
\end{subequations}
where $ \mathcal{P}_x^{SS} \equiv 2S^z_{x}S^z_{x+1}\! +\! 1/2$ is $1$ if
$S^z_{x}=S^z_{x+1}$ and zero otherwise, and 
$\sum'$ denotes a sum only
on sites that are outside of the brackets ($i \notin [x\!-\!l,x\!+\!r]$).
We have thus obtained that, using the mapping, the coupling between the 
hole and spin degrees of freedom seems to be
 limited to the interval between the brackets.
Therefore, it is natural to construct a variational wavefunction by starting
with a complete decoupling between the hole and the stripe, and project out the
``illegal'' states, in the spirit of the Gutzwiller projection.

\section{Motion of a hole on a stripe}
In order to examine the motion of a hole on a stripe we
 find it instructive to represent the various states of the stripe by a
\emph{state graph}~\cite{ng_stripe}, which is defined such that each 
node represents a state
and each line represents an allowed hop between states.
In order for the hole and brackets to shift by one position, in this model, 
 three individual hops are required (see, e.g., fig.~\ref{fig:main_path}). 
Generally, it is clear that the allowed hole and bracket moves depend on the 
background stripe configuration. 
The simplest possible motion is when each of the brackets does not stray 
more than one position away from the hole, 
thereby eliminating any dependence on the stripe. 
This motion can be represented in the state space by the solid 
lines in fig.~\ref{fig:states}, and we refer to it as the \emph{``main path''}.
Depending on the local tilt in the stripe (runs of spins in the same
direction), the available states are increased.
In the extreme case, if the stripe is a $45^\circ$ diagonal, then
the hole and brackets can be treated as three non-interacting fermions and 
their energy is lower than in the untilted case. 
This implies that the hole prefers the stripe to be locally tilted around it. 
However, the spin chain loses energy by having a region where all of 
the spins are the same, since no spin exchange is possible there. 
We would expect the interplay between these two competing effects to result 
in a small tilted region (``kink'') around the hole. 
This has indeed been inferred from ED of ref.~\cite{ng_stripe}, 
and is one of the phenomena that we look for in our solution for this model.
\begin{figure}
\onefigure[scale=0.67]{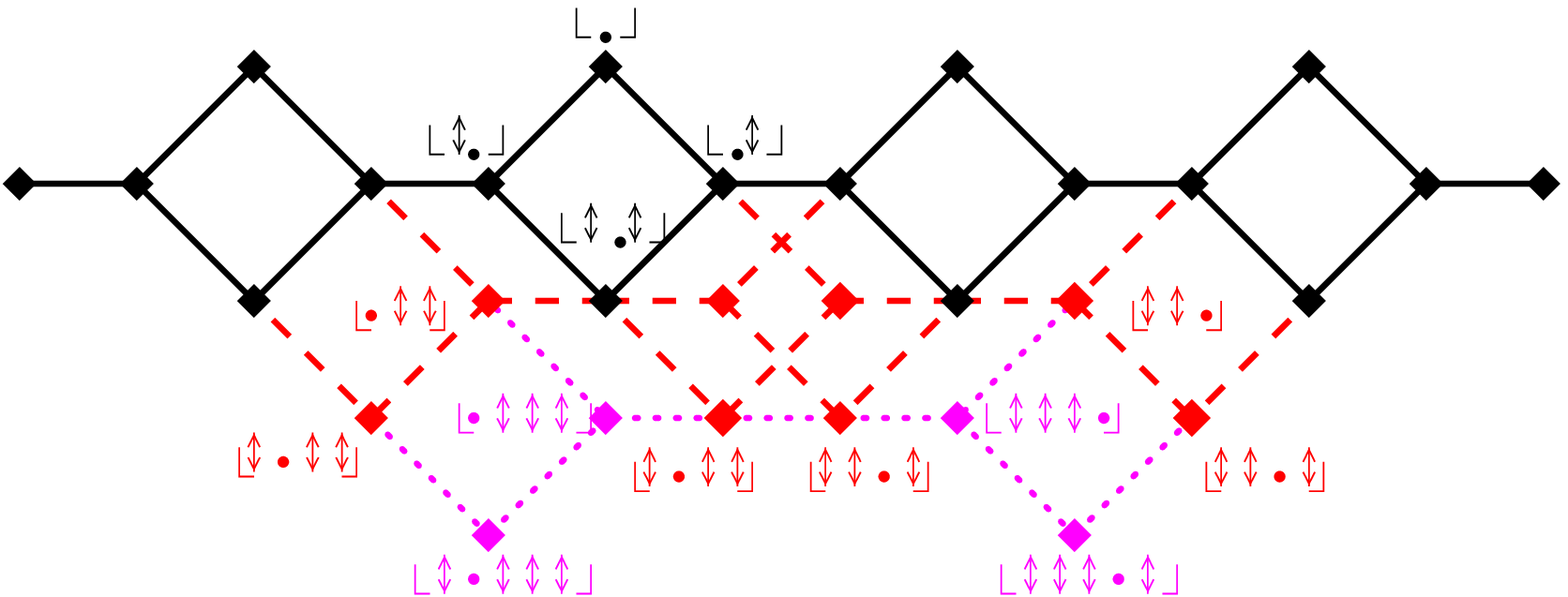}
\caption{State space representation of hole motion for a particular 
background stripe/spin configuration. 
The $\updownarrow$ symbol represents an unspecified spin 
($\uparrow$ or $\downarrow$).
The solid lines represent the ``main path'', in which the hole motion is decoupled from its environment. The dashed and dotted lines represent the additional available hops when there is a run of two or three spins in the same 
direction, respectively. 
If we read this from left to right, a horizontal line represents one hop of the hole
to the right, a downward(upward) diagonal line represents hopping of the 
right(left) bracket. }
\label{fig:states}
\end{figure}

Examining the motion of a single hole on a stripe, we find that any sequence
of moves that returns the system to its original configuration involves 
an even permutation, 
except for hole hopping around odd boundary conditions~\cite{bose_stats}.
This implies that for even or infinite $L_x$, the boson and fermion spectra 
are identical.

\section{\label{sec:var}Variational wavefunction}
In order to calculate the hole spectrum variationally, 
we introduce a projection operator $\proj_{lr}^{\alpha \beta}(x)$ acting on 
states of a spin chain of length $L_x'$,
to the sub-space where all of the $l$ spins at sites $[x\!-\!l\!+\!1,x]$ are 
in direction $\alpha$,
 and $r$ spins at positions $[x\!+\!1,x\!+\!r]$ are in direction $\beta$
($\alpha, \beta \in \{\downarrow,\uparrow\}$). $\proj_{lr}^{\alpha \beta}(x)$
annihilates states that are not in this subspace.
We define an orthonormal basis set:
\begin{equation}
\chi_{lr}^{\alpha \beta}(x) =
\frac{1}{\mathcal{N}_{lr}^{\alpha \beta}}
|x,l,r\rangle \proj_{lr}^{\alpha \beta}(x) | \Phi_F \rangle\,,
\label{eq:basis}
\end{equation}
where
$| \Phi_F \rangle$ is ground state of an unrestricted spin-$1/2$ ring of length $L'_x$;
 $\mathcal{N}_{lr}^{\alpha \beta}=\sqrt{\langle \proj_{lr}^{\alpha \beta}\rangle_F}$ is a normalization factor,
 where $\langle \proj_{lr}^{\alpha \beta} \rangle_F \equiv  
\langle \Phi_F | \proj_{lr}^{\alpha \beta} (x) | \Phi_F \rangle$ is 
independent of $x$.

Now we want to calculate $\langle \chi_{lr}^{\alpha \beta}(x) 
|\ham_\mathrm{hole}| \chi_{l'r'}^{\alpha' \beta'} (x) \rangle $. 
The matrix elements for $\ham_h$, $\ham_r$, $\ham_l$ are straightforward,
but those of $\ham'_{sp}$ are a little harder to calculate. 
Since $\ham'_{sp}$ does not change the hole+brackets state $|x,l,r\rangle$, 
we only need matrix elements between pairs of states with the same 
$x$,$l$,$r$, i.e. $\langle \chi_{l r}^{\alpha \beta}(x)| \ham'_{sp}| \chi_{l r}^{\alpha' \beta'}(x) \rangle$. 
These can be calculated using the fact that $\ham'_{sp}$ 
commutes with $\proj_{l r}^{\alpha \beta}(x)$
\begin{eqnarray}
  \langle \chi_{lr}^{\alpha \beta}(x) |\ham'_{sp} | \chi_{lr}^{\alpha' \beta'}(x) \rangle   &=&
\frac{1}{\mathcal{N}_{lr}^{\alpha \beta} \mathcal{N}_{lr}^{\alpha' \beta'}}
 \langle \proj_{l r}^{\alpha \beta}(x)  
{ \sum_i}' \ham_{ex}^i 
\proj_{l r}^{\alpha' \beta'}(x) \rangle_F  \nonumber \\ &=&   
\frac{1}{\langle \proj_{l r}^{\alpha \beta} \rangle_F} 
\tilde{\delta}^l_{\alpha,\alpha'} \tilde{\delta}^r_{\beta,\beta'} 
\langle \proj_{l r}^{\alpha \beta}(0) 
{ \sum_i}' \ham_{ex}^i  \rangle_F \nonumber \\&=&   
\tilde{\delta}^l_{\alpha,\alpha'} \tilde{\delta}^r_{\beta,\beta'}  \left[ \epsilon_0^{\scriptscriptstyle{(L'_x)}} -
\frac{\langle \proj_{l r}^{\alpha \beta}(0) 
( \ham_{ex}^{x-l}+\ham_{ex}^x+\ham_{ex}^{x+r}) \rangle_F}
{\langle \proj_{l r}^{\alpha \beta} \rangle_F }
\right] \,,
\end{eqnarray}
which is independent of $x$. We used:
(i) $[\ham_{ex}^i,\proj_{l r}^{\alpha \beta}(x)]=0$ for 
$i \notin [x\!-\!l,x\!+\!r]$ (modulo $L'_x$).
(ii) $\proj_{l r}^{\alpha \beta}(x) \ham_{ex}^i =0$ for 
$i\in (x\!-\!l,x)$ or $i\in (x, x \!+\!r)$.
Note that when $l\!=\!0$, $\alpha$ is undefined, so in order to generalize the 
notations, we arbitrarily set $\alpha=\downarrow$ when $l\!=\!0$ 
($\proj_{l=0,r}^{\uparrow \beta}\!=\! 0$),
 and similarly for $r,\beta$.
 We defined 
$\tilde{\delta}^l_{\alpha,\gamma} \equiv  \delta_{\alpha,\gamma}$ if 
$l>0$ and $\tilde{\delta}^l_{\alpha,\gamma} = 0$ otherwise.
In order to calculate the energy spectrum using the variational wavefunction,
we Fourier transform  $\chi_{lr}^{\alpha \beta}(x)$ and find  
$\langle \chi_{lr}^{\alpha \beta}(k) |\ham_{hole} 
| \chi_{lr}^{\alpha' \beta'}(k) \rangle$ in a straightforward manner.

\section{\label{sec:res}Results}
The results presented here were obtained for 
a subset of the variational basis set with $l\!+\!r \leq 10$ for
our calculations.
If one does not take advantage of mirror symmetries between states, 
the Hamiltonian is a $221 \! \times \! 221$ matrix.
We find the ground state and ground state energy of this Hamiltonian 
as a function of $L_x$, $b$, and $k$. 
Increasing the variational basis set to up to $l\!+\!r \leq \!13$ 
results in a relative change of less than $0.1\%$ in $\Delta$ for the
maximum overall tilt presented here,
and less than $10^{-14}$ for zero tilt.
In our calculation, we take into account the fermion statistics only
by adding a phase to hole hops in (finite) odd boundary conditions. 
This is justified because these are the only moves that can induce an odd
cyclic permutation of the particles~\cite{bose_stats}.

\begin{figure}
\twofigures[scale=0.35]{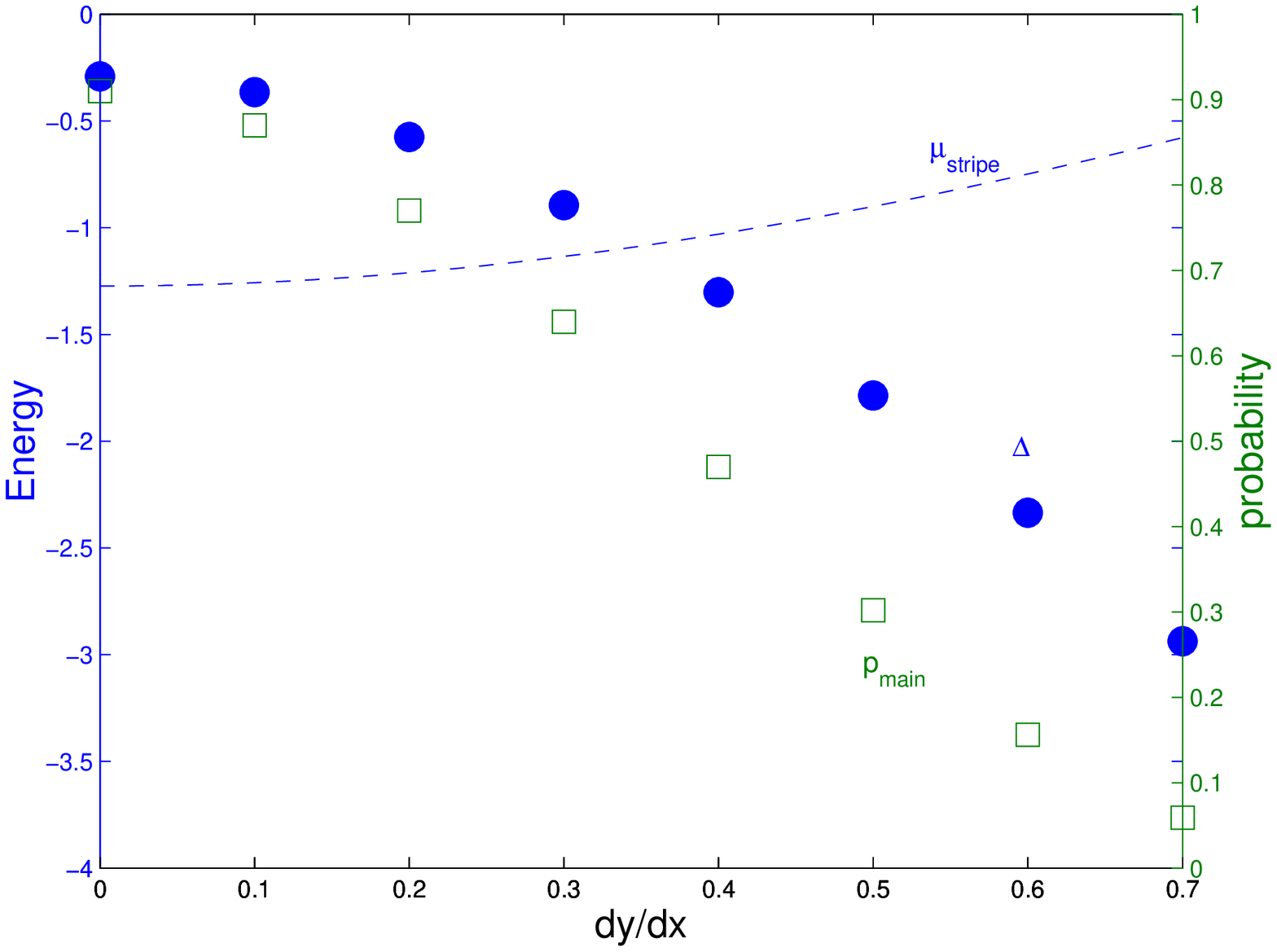}{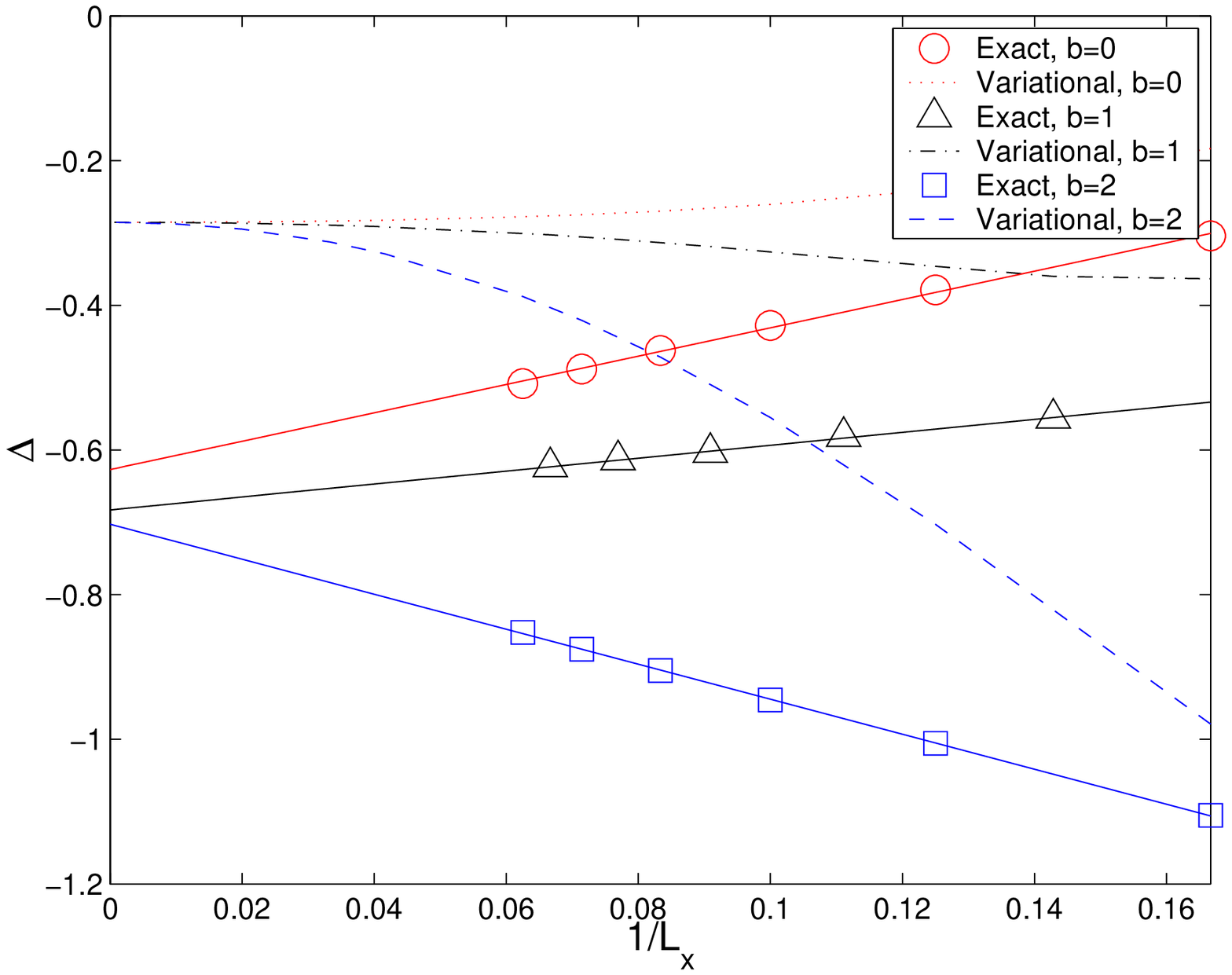}
\caption{Circles: Ground state energy difference
 $\Delta\!\equiv\! E_{hole}\!-\!E_{stripe}$
 as a function of overall
stripe tilt, for infinite $L_x$.
 The dashed line represents 
the chemical potential $\mu_{stripe}$. 
($\mu_{stripe}\!<\!\Delta$ is required for stripe stability).
Squares: Main path probability $P_{main}$.}
\label{fig:del_p}
\caption{Energy difference $\Delta$ obtained in
 ED and variational calculations, for finite system sizes.
Results are shown
for horizontal stripes ($b\!=\!0$), as well as 
tilted boundary conditions ($b\!=\!1,\!2$). The solid lines are linear 
interpolations of the ED data. Note that although 
$\Delta$ decreases as the tilt increases,
$E_{hole}$ generally increases with tilt.}
\label{fig:energy}
\end{figure}

Figure~\ref{fig:del_p} shows the ground state energy difference 
$\Delta$,
as a function of the overall stripe tilt.
We obtain  $\Delta=-0.29$ for an (infinite)
untilted stripe.
As the stripe tilt is increased, both the energy of the undoped stripe, 
$E_{stripe}$, and $E_{hole}$ increase,
 because stripe fluctuations are reduced.
 However, the difference $\Delta$ decreases, because the hole's kinetic energy
is enhanced by the additional tilt.
Figure~\ref{fig:energy} shows $\Delta$ for 
finite system sizes, compared to respective ED results of 
ref.~\cite{ng_stripe}, and their extrapolation to $L_x \rightarrow \infty$. 
The variational energy is, of course,
 an upper bound to the true energy of the system.
In the ED of ref.~\cite{ng_stripe}, $E_{hole}$ was minimal 
for stripes with slightly tilted boundary conditions ($b\!=\!2$ for even $L_x$, 
$b\!=\!3$ for odd $L_x$). This suggested that the hole
tends to bind to a kink in the stripe in order to increase 
its kinetic energy, at the expense of stripe fluctuations, forming a polaron.
In our variational calculation, we observed the same effect for
 small system sizes (up to $L_x\! \approx \!10$), 
however, for larger systems, the minimum energy is for $b\!=\!0(1)$ 
for even(odd) $L_x$.
This indicates that the preference for
tilted boundary conditions, as observed in ED,
may be only a finite size effect.
We also calculated the hole dispersion. On an untilted stripe, we find
an effective mass $m_h^*\!=\!6.43m^*$, where $m^*\!=0\!.5$ is the 
mass of a non-interacting particle hopping on the lattice. 
This agrees remarkably well with ED, 
which gives $m_h^*\! \approx\! 6.7m^*$ \cite{ng_stripe}.
If a hole would be forced to move only on the ``main path'', its effective
mass would be $7.35m^*$.
When the stripe has an overall tilt, the hole mass is reduced 
significantly, e.g. at $dy/dx\!=\!0.5$, $m_h^*\!=\!4.57m^*$.

A good measure of the coupling between the hole and the stripe is
$p_{main}$, the ground state probability of being in one of the ``main path''
basis states. 
In the variational ground state for an untilted stripe, we find
$p_{main}\!=\!0.91$, i.e. the hole tends to be decoupled from the stripe 
configuration.  This probability is reduced as the stripe tilt 
increases and it is easier for the brackets to move from the hole
(see fig.~\ref{fig:del_p}). This trend is supported by 
ED calculations, but the variational values for $p_{main}$ are higher by up
to $0.1$.
Comparison of the variational and ED ground states,
reveals that the our calculation does a good job of
qualitatively capturing the composition of the exact ground state
from the basis set (e.g. table~\ref{tab:g_state}), but
it overestimates the weight of states with low $l\!+\!r$.
\begin{table}
\caption{\label{tab:g_state} 
Variational and ED ground state probabilities
for $(10,0)\!\times\!(0,7)$.
All other states, except those related by symmetry to the ones presented
have probability of less than 1\%. } 
\begin{largetabular}{|l|ccccc|}
\hline
Basis State & $\chi_{00}^{\down \down}$ &  $\chi_{01}^{\down \up}$ &
  $\chi_{11}^{\down \up}$ &  $\chi_{11}^{\down \down}$ & 
 $\chi_{02}^{\down \down}$ \\
\hline
variational  &  23.4\%  & 13.7\% & 4.3\% & 2.9\% & 1.4\% \\
ED  &  19.2\%  & 12.3\% & 4.7\% & 3.6\% & 1.8\% \\
\hline
\end{largetabular}
\end{table}

In summary, we calculated the energy and the ground state for
a single hole on a stripe, using a mapping to one dimension 
and a variational wavefunction constructed to decouple the hole 
and stripe degrees of freedom.
We did not find evidence that the hole binds to a kink in the stripe, in the
untilted case, for $L_x\! \gtrsim\! 10$.

\acknowledgements
We thank N. G. Zhang for use of his ED computer program.
C. L. H. thanks S. Petrosyan for discussions.
Support for this work was provided by NSF grant DMR-9981744.


\begin{thebibliography}{0}
\bibitem{flux_footnote}
In this mapping,
we did not consider the possible
overall motion of the stripe in the $y$ direction. 
However, this only adds a phase to the wavefunction, which
does not affect our results~\cite{ng_stripe}.
\end{thebibliography}
\end{document}